\documentclass[prc,aps,floats,showpacs,twoside,preprintnumbers, superscriptaddress,floatfix,twocolumn]{revtex4}
\usepackage{graphicx,pstricks,epsfig,rotating,amsmath,amssymb,color}

\newcommand{\bea}{\begin{eqnarray}}
\newcommand{\eea}{\end{eqnarray}}
\newcommand{\apgt} {\ {\raise-.5ex\hbox{$\buildrel>\over\sim$}}\ }

\graphicspath{{EsymPaperFigs/}}

\begin{document}
\title{High-mass twin stars with a multi-polytrope EoS}
\date{\today}

\author{D.~E.~Alvarez-Castillo}
\email{alvarez@theor.jinr.ru}
\affiliation{
Bogoliubov  Laboratory of Theoretical Physics,
Joint Institute for Nuclear Research,\\
Joliot-Curie str. 6, 141980 Dubna, Russia}
\affiliation{
GSI Helmholtzzentrum f\"ur Schwerionenforschung GmbH, Planckstra{\ss}e 1, 64291 Darmstadt, Germany}
\affiliation{Instituto de F\'{\i}sica,
Universidad Aut\'onoma de San Luis Potos\'{\i}\\
Av. Manuel Nava 6, San Luis Potos\'{\i}, S.L.P. 78290, M\'exico}

\author{D.~B.~Blaschke}
\email{blaschke@ift.uni.wroc.pl}
\affiliation{
Bogoliubov  Laboratory of Theoretical Physics,
Joint Institute for Nuclear Research,\\
Joliot-Curie str. 6, 141980 Dubna,
Russia}
\affiliation{
Institute for Theoretical Physics,
University of Wroc{\l}aw,
Max Born Pl. 9,
50-204 Wroc{\l}aw, Poland}
\affiliation{National Research Nuclear University (MEPhI),
Kashirskoe Shosse 31,
115409 Moscow, Russia}


\begin{abstract}
We show that in the three-polytrope model of Hebeler et al. \cite{Hebeler:2013nza} for the neutron star equation of state at supersaturation densities a third family of compact stars can be obtained which confirms the possibility of high-mass twin stars that have coincident masses 
$M_1=M_2\approx 2~M_\odot$ and significantly different radii $|R_1-R_2|>2-3 $ km.
We show that the causality constraint puts severe limitations on the maximum mass of the third family sequence which can be relaxed when this scheme is extended to four polytropes thus mimicking a 
realistic high-density matter EoS.
\end{abstract}

\maketitle

\section{Introduction}

Compact stars (CS) represent one possible endpoint of stellar evolution with conditions of very high density in their interiors for which the equation of state (EoS) is currently unknown.
In particular, the composition of CS interiors is puzzling: are they composed of exotic forms of matter like hyperonic matter or phases of quark matter which even may be color superconducting \cite{EPJA:2016exotic}?  	
In this connection arises the question for the order of the transition to these exotic phases.
If it was to be a first-order transition, this would imply the existence of at least one critical endpoint in the QCD phase diagram, since we know from lattice QCD simulations that at vanishing baryon density and finite temperatures the QCD transition is a crossover \cite{Bazavov:2014pvz}.

In 2013, we have suggested \cite{Alvarez-Castillo:2013cxa,Blaschke:2013ana} that there is a possibility to decide this question by CS observations
following the classification scheme \cite{Alford:2013aca} of  mass-radius  ($M-R$) diagrams for hybrid star EoS in dependence on the characteristic features of the transition, 
the jump in energy density $\Delta\varepsilon$ and the critical pressure $P_{\rm crit}$ at the onset of the transition.
For instance, if the $M-R$ diagram of CS would feature the so-called \textit{third family} branch that is separated from the second family branch of neutron stars (NS) as purely hadronic CS by a sequence of unstable configurations, this would indicate the EoS of CS matter has a phase transition with a sufficiently large $\Delta\varepsilon$.
Since the conversion of the core matter to a phase with higher density at the same pressure entails a compactification of the star accompanied with a release of gravitational binding energy and therefore a lowering of the gravitational mass, the second and third family branches do overlap in a certain mass window.
This situation is called the mass twin phenomenon: two stars of the same mass would be located in the second and third family branches respectively, having different radii and internal composition.
If this occurs at a high mass of $\sim 2M_\odot$ one speaks of the high-mass twin (HMT) scenario. 
The most prominent example being due to the hadron-to-quark-matter phase transition
resulting in hybrid CS composed of a hadronic mantle and a quark matter core, see \cite{Benic:2014jia, Alvarez-Castillo:2016wqj} and references therein.   
The HMT scenario is in principle accessible to observational verification, e.g., by satellite missions like NICER \cite{nicer} or ground-based programs as SKA \cite{ska}. 
All it takes is to measure radii of NS with sufficiently similar high masses such as PSR J0348+0432 with $M=2.01\pm 0.04~M_\odot$ \cite{Antoniadis:2013pzd} and PSR J1614-2230 with $M=1.928 \pm 0.017~M_\odot$ \cite{Demorest:2010bx,Fonseca:2016tux}, and to find out that their radii are significantly different.

In fact, there are a few approaches in the literature on description of CS twins. First realizations include~\cite{Glendenning:1998ag} using a Relativistic Mean Field model for the hadronic together with a bag model for the quark matter phase as
well as~\cite{Schertler:1998cs} where strange quark matter is considered instead. However, due to the soft hadronic equations of state used there, the resulting twin stars cannot reach the $M=2.01M_\odot$, a measurement at that time unknown. 
On the other hand,~\cite{Alford:2013aca,Alford:2015dpa} feature a fixed hadronic EoS together with a constant speed of light quark matter parameterization and present a phase diagram for all masses and different mass-radius topologies,
namely connected and disconnected branches. Another alternative
description is reported in~\cite{Zacchi:2016tjw} where a SU(3) chiral-meson model~\cite{Zacchi:2015oma} together with a bag constant quark matter
model is implemented and leads to very large CS radii incompatible with the Hebeler constraints. Last but not least, a few other realizations of HMTs have been reviewed in~\cite{Alvarez-Castillo:2016wqj} where the connection with heavy ion collision experiments
has been underlined.

Furthermore, as pointed out in~\cite{Blaschke:2015uva}, the HMT phenomenon is of great relevance for the study of the NS EoS not only because it can provide evidence for a first order phase transition and thus for the very existence of a critical endpoint in the QCD diagram, but also because it provides a resolution to several issues: 
the hyperon puzzle~\cite{Weissenborn:2011kb}, the reconfinement problem and the masquerade case
(see~\cite{Blaschke:2015uva} and references therein). 
In addition, the HMT may be discussed in the context of explaining the origin of fast radio bursts~\cite{Champion:2015pmj} as possible intermediate metastable states due to a sudden change in the internal structure of a fastly rotating supramassive neutron star \cite{Alvarez-Castillo:2015dqa,Bejger:2016emu} created, e.g., in a NS merger event before its final collapse to a black hole 
\cite{Falcke:2013xpa}.

The purpose of this work is to point out that the HMT case is not the result of the construction of a rather exotic case of an EoS but may be obtained even within the rather conservative scheme of Hebeler et al.~\cite{Hebeler:2013nza}. 
It consists of a multi-polytrope description of the NS EoS~\cite{Read:2008iy} in line with constraints derived from a chiral effective field theory describing nuclear few- and many-particle systems at densities up to nuclear saturation.

\section{Piecewise polytrope EoS with a first-order phase transition}
We would like to investigate the question whether in the scheme of Hebeler et al. \cite{Hebeler:2013nza}
with a piecewise polytrope EoS at supersaturation densities it would be possible to describe the HMT 
phenomenon.
To this end one should define one of the polytropes as a constant pressure region with 
$P=P_{\rm crit}$ with a jump in energy density $\Delta\varepsilon$ due to a first order phase transition 
that would fulfil the Seidov constraint 
\cite{Seidov:1971} 
\bea
\label{seidov}
\frac{\Delta\varepsilon}{\varepsilon_{\rm crit}} \ge \frac{1}{2} 
+ \frac{3}{2} \frac{P_{\rm crit} }{\varepsilon_{\rm crit}} 
\eea
for the occurrence of an instability in the M-R relation of compact stars.
Such a sequence of unstable configurations is precondition for a disconnected (third family) branch of stable hybrid stars that would furthermore require a sufficiently stiff high density EoS to allow for a maximum mass fulfilling the constraint from the measurement of the mass 
$M=2.01\pm 0.04~M_\odot$  for the pulsar PSR J0348+0432 \cite{Antoniadis:2013pzd}.
According to \cite{Hebeler:2013nza}, the supersaturation density region is split into three regions
\bea
i=1&:& n_1 \le n \le n_{12}\nonumber\\
i=2&:& n_{12} \le n \le n_{23}\\
i=3&:& n \ge n_{23}\nonumber~,
\eea
where $n_1 = 1.1~n_0$ is just above the nuclear saturation density $n_0=0.15$ fm$^{-3}$ and the polytrope EoS pieces fulfill 
\bea
\label{polytrope}
P_i(n)=\kappa_i n^{\Gamma_i}
\eea
in the corresponding regions.
For our setting of the problem, in all our EoS models the first region is taken from the Hebeler et al. paper~\cite{Hebeler:2013nza} and corresponds to a polytrope fit to the stiffest EoS ($n>1.1~n_0$) 
of their table V together with an intermediate homogeneous phase in $\beta-$ equilibrium 
($0.5~n_0<n< 1.1~n_0$), presented in their section III, and the BPS EoS for the outer NS crust ($n<0.5~n_0$) of their table VII. 
Therefore, the resulting fit gives polytrope parameter values for this density region of~$\Gamma_1=4.92$ and $\kappa_1=17906.60$~MeV$\cdot$fm$^{3(\Gamma_1-1)}$. 
Furthermore, the region $i=2$ shall correspond to the phase coexistence region of the
first order phase transition with constant pressure, so that $\Gamma_2=0$ and 
$P_2 = \kappa_2=P_{\rm crit}$.
The boundaries of this region shall be obtained from a Maxwell construction~\cite{Bhattacharyya:2009fg} which requires the pressure as a function of the chemical potential $\mu$. 
To facilitate this construction for a pair of polytropes at zero temperature, we utilize the formulae given in  the Appendix of Ref.~\cite{Zdunik:2005kh},
\bea
P(n)&=&n^2 \frac{d(\varepsilon(n)/n)}{dn},\\
\varepsilon(n)/n &=& \int dn \, \frac{P(n)}{n^2} = \int dn\, \kappa n^{\Gamma-2} 
\nonumber\\
&=& \frac{\kappa\, n^{\Gamma-1}}{\Gamma-1} + C,\\
\mu(n)&=& \frac{P(n) + \varepsilon(n)}{n} =  \frac{\kappa\, \Gamma}{\Gamma-1} n^{\Gamma-1} + m_0, 
\label{5}
\eea
where the integration constant $C$ is fixed by the condition that $\varepsilon(n\to 0)=m_0\, n$.   
Now we have for the polytrope EoS
\bea
n(\mu)&=&\left[ (\mu-m_0)\frac{\Gamma-1}{\kappa \Gamma}\right]^{1/(\Gamma-1)},
\eea
so that the pressure as a function of the chemical potential for the polytrope EoS (\ref{polytrope}) is
\bea
\label{P-mu}
P(\mu)&=&\kappa \left[ (\mu-m_0)\frac{\Gamma-1}{\kappa \Gamma}\right]^{\Gamma/(\Gamma-1)}~.
\eea
With EoS in the form (\ref{P-mu}) one can perform the Maxwell construction of a first-order phase transition.

\section{High mass twins from multi-polytrope equations of state}

Now we can apply these general relations to the case of a transition from nuclear matter in the region 1 
(with $m_{0,1}$ being the nucleon mass)  to high density matter in region 3. 
That may correspond, e.g., to hyperon matter 
or quark matter.
From the Maxwell construction 
\bea
P_1(\mu_{\rm crit})&=&P_3(\mu_{\rm crit})=P_{\rm crit}\\
\mu_{\rm crit}&=&\mu_1(n_{12})=\mu_3(n_{23})
\eea
follow the two conditions
\bea
\label{10}
\kappa_3&=&\kappa_1\, {n_{12}^{\Gamma_1}}/{n_{23}^{\Gamma_3}},\\
P_{\rm crit} &=& \left(m_{0,1} - m_{0,3} \right) 
\left[ \frac{\Gamma_1}{n_{12}(\Gamma_1-1)} -\frac{\Gamma_3}{n_{23}(\Gamma_3-1)} \right]^{-1}
\nonumber\\
&=&\kappa_1\, n_{12}^{\Gamma_1}.
\label{11}
\eea

The above equations~(\ref{10}) and~(\ref{11}), allow for determination of $\kappa_3$ and $m_{0,3}$ once the values of $n_{12}$, $n_{23}$, $m_{0,1}$ and $\Gamma_3$ are fixed.
In order to fulfill the compact star mass constraint, we may demand that upon solving the corresponding compact star sequence, 
the mass at the onset of the transition fulfills $M(n_{12})\simeq 2\, M_\odot$ which fixes the value $n_{12}$, allowing to determine $P_{\rm crit}$ and $\mu_{\rm crit}$ from the Maxwell construction. 
Moreover, since the EoS just above nuclear saturation is fixed according to the stiff limit of Hebeler et al. 
\cite{Hebeler:2013nza}, the constants $\Gamma_1$ and $\kappa_1$ are also fixed.
So we are then left with the three equations (\ref{10}) and (\ref{11}) and (\ref{seidov}) for the
four unknowns $\kappa_3$, $\Gamma_3$ and $n_{23}$ and $m_{0,3}$. 
We fix $n_{23}$ such that the equality sign holds in the Seidov criterion (\ref{seidov}).  
Therefore, in order to close the system, we dial $\Gamma_3$ as a free parameter. 
Its maximal value is determined so that the speed of sound shall not exceed the speed of light up to the densitiy values reached in the very center of the maximum mass star configurations.

With the above scheme, we are thus able to compute the EoS of hybrid compact stars.
Next, we obtain the corresponding star sequences by solving the Tolman-Volkoff-Oppenheimer (TOV) equations describing a static, non-rotating, spherically symmetric star
\cite{Tolman:1939jz,Oppenheimer:1939ne}
\bea
\frac{dP( r)}{dr}&=& 
-\frac{G\left(\varepsilon( r)+P( r)\right)
	\left(M( r)+ 4\pi r^3 P( r)\right)}{r\left(r- 2GM( r)\right)},\\
\frac{dM( r)}{dr}&=& 4\pi r^2 \varepsilon( r),
\eea
with $P(r=R)=0$ and $ P_c= P(r=0)$ as boundary conditions for a star with mass $M$ and radius $R$. 
The complete NS sequence is determined by increasing the chosen central pressure $P_c$ up to a maximum mass. 

\begin{widetext}
	
	\begin{figure*}[!htb]
		\includegraphics[width=0.9\textwidth, angle=0]{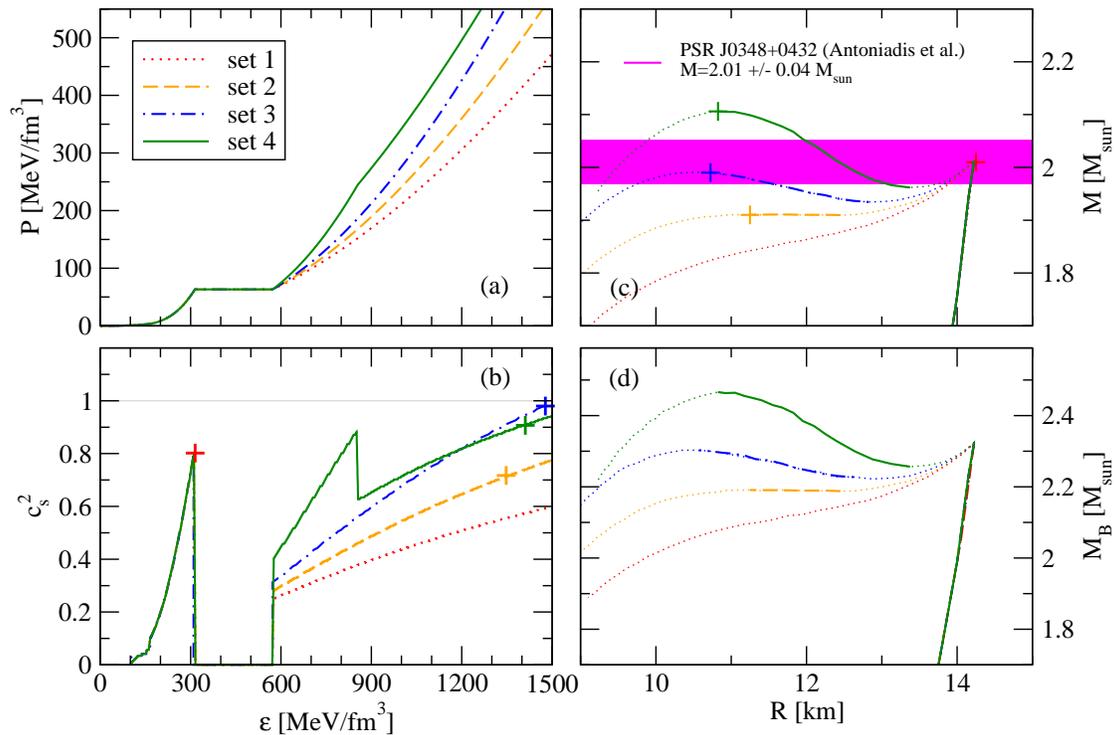}
		\caption{(Color online) 
			EoS ((a) - P vs. $\varepsilon$; (b) - $c_s^2$ vs. $\varepsilon$) and sequences of compact stars 
			(({c}) - $M$ vs. $R$; (d) - $M_B$ vs. $R$) for sets 1-4 of table \ref{param-123}.
			The EoS have the same onset density and density jump of the phase transition, but different stiffness of the high density (quark matter) phase.
			The plus symbols denote values for the maximum mass configurations. 
			\label{MP-same-onset}}
	\end{figure*}
	
\end{widetext}

The enclosed baryonic mass is obtained by integrating
\bea
\frac{d N_B( r)}{dr}&=& 4\pi r^2 \left(1-\frac{2GM( r)}{r}\right)^{-1/2}n( r)~.
\eea

It plays an important role in the description of the dynamics of CS evolution scenarios 
that conserve the baryonic mass, like spinning down into the hybrid twin configuration
\cite{Bejger:2016emu}. 
In our numerical calculations, we vary the values of the polytrope index $\Gamma_3$ in order to 
discuss the effect of the stiffness of the high-density EoS 
using the sets 1-4 of model parameters 
given in Table~\ref{param-123}.
\begin{table}[!htb]
	\centering
	\caption{Parameter values for sets 1, 2, 3 and 4. The EoS in this set share the following properties:
		$P_{\rm crit}=63.177$ MeV fm$^{-3}$, $\varepsilon_{\rm crit}=318.26$ MeV fm$^{-3}$, 
		$\Delta \varepsilon=253.89$ MeV fm$^{-3}$. The second polytrope with $P_2=P_{\rm crit}$ and
		$\Gamma_2=0$ lies between the densities $n_{12}=0.32$ fm$^{-3}$ and $n_{23}=0.53$ fm$^{-3}$.
		For set 4 the high-density region has been divided in two polytrope branches: set 4a for densities
		$n_{23}\le n \le n_{34}$ and set 4b for densities $n\ge n_{34}=0.75$ fm$^{-3}$.}
	\label{param-123}
	\begin{tabular}{l|ccc|ccc}
		\hline \hline
		&$\Gamma_3$ & $\kappa_3$& $m_{0,3}$ &$M_{\rm max}^{NS}$  & $M_{\rm max}^{HS}$ & $M_{\rm min}^{HS}$\\		
		&& [MeV fm$^{3(\Gamma_3-1)}$]& [MeV] & [M$_{\odot}$] & [M$_{\odot}$] & [M$_{\odot}$] \\			
		\hline
		set 1&2.50 &302.56& 991.75 &2.01 & - & - \\
		set 2&2.80 &365.12& 1004.88 &2.01 & 1.910 & 1.909 \\
		set 3&3.12 &447.16& 1014.87 & 2.01 & 1.991 & 1.934 \\
		set 4a&4.00 &774.375& 1031.815 & & &  \\
		set 4b&2.80 &548.309& 958.553 & 2.01 & 2.106 & 1.961 \\
		\hline \hline
	\end{tabular}
\end{table}

\begin{figure}[!ht]
	\centering
	\includegraphics[width=1.0\hsize]{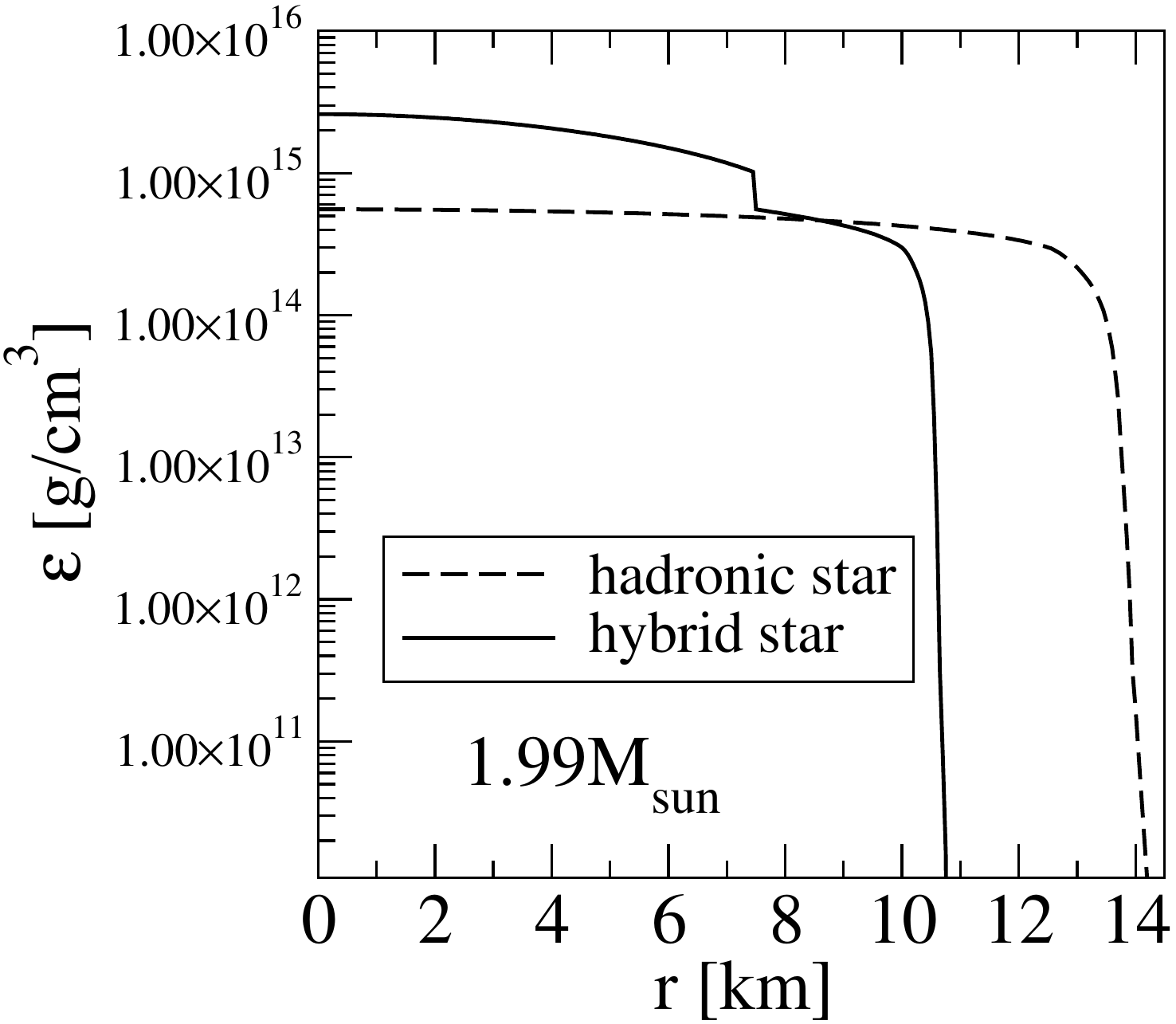}
	\caption{
		High mass twins internal energy density profile. The dashed curve corresponds to the pure hadronic star whose radius exceeds the hybrid star by about 3 kilometers.
		\label{internal_profile}}
\end{figure}

	The hybrid star EoS  corresponding to sets 1-4
share the same hadronic branch, onset and density jump at the phase transition, see the panel (a) of Fig.~\ref{MP-same-onset}. 
On panel (b) of Fig.~\ref{MP-same-onset} we demonstrate that the causality constraint is fulfilled for all these sets.
Increasing the quark matter stiffness through the parameter $\Gamma_3$ results in an increase of the maximum mass on the hybrid star branch.
The EoS for set 3 reaches the causality limit just at the maximum mass of the hybrid star sequence,
which is denoted by the plus symbols on panels (b) and ({c}) of that figure. 
It shows that a third family of stable hybrid stars can be obtained within the multi-polytrope scheme of Hebeler et al.~\cite{Hebeler:2013nza}
using just three polytropes.
This is the main result of this paper.
The hybrid star branch  is more compact than the
purely hadronic one by about $2 - 3$ km. 
In Fig.~\ref{internal_profile} we illustrate the case of HMT stars by showing for set 3 the energy density profiles of two stars with the same mass of $1.99~M_\odot$, one from the hadronic branch and one from the hybrid star branch. The latter has an extended quark matter core and is more compact than its hadronic twin by about 3 km.  
This is a potentially observable effect!

Another important result is obtained when going beyond the three-polytrope scheme of  Hebeler et al.~\cite{Hebeler:2013nza} by introducing a fourth polytrope at high densities 
in order to prevent a causality breach.
With set 4 of our study we achieve an increase in the hybrid star maximum mass which now even reaches $\sim 2.1~M_\odot$.

Actually, the physical motivation for adopting a change in the polytrope index $\Gamma_3$ of the 
high-density matter comes from the fact that there can be a sequence of phase transitions in CS matter 
at high densities \cite{Blaschke:2008br}, which may even lead to the occurrence of a fourth family 
of CS \cite{Alford:2017qgh}.
Recently we could show that also a fifth family solution is possible for the case of three sequential transitions in CS matter \cite{5thfamily}, as discussed in \cite{Blaschke:2008br}.

Lowering $\Gamma_3$ we obtain a value for which barely a stable hybrid star sequence can be obtained (set 2), and lowering $\Gamma_3$ further (set 1) no stable hybrid stars are possible. 
The four parameter sets for which the EoS and compact star sequences are illustrated in Fig.~\ref{MP-same-onset} are given in Table~\ref{param-123}. 
Note that for set 4 the high-density region has been divided in two polytrope branches: set 4a for densities
$n_{23}\le n \le n_{34}$ and set 4b for densities $n\ge n_{34}=0.75$ fm$^{-3}$.  

\begin{figure}[!h]
	\includegraphics[width=1.0\hsize]{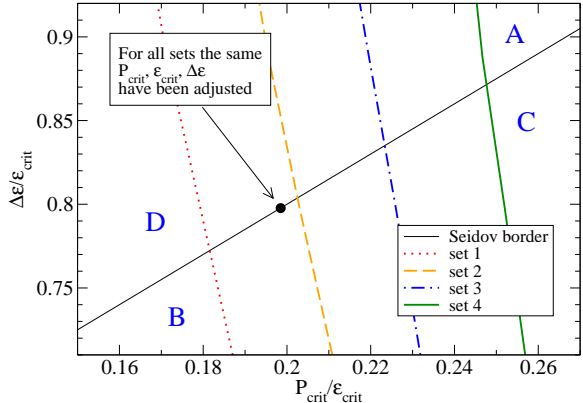}		
	\caption{"Phase diagram" of hybrid star sequences following Ref.~\cite{Alford:2013aca} for the parameter sets 1-4 of this work. For details, see text.
		\label{fig:AHP-phases}}
\end{figure}

It is interesting to discuss the present parametrizations of the multi-polytrope EoS in the classification scheme of Ref.~\cite{Alford:2013aca} where a phase diagram for hybrid CS sequences with a phase transition
has been introduced that is spanned by the plane of $\Delta \varepsilon$ and $P_{\rm crit}$, both measured in units of $\varepsilon_{\rm crit}$, see Fig.~\ref{fig:AHP-phases}.
By construction all our parametrizations have the same $P_{\rm crit}$, $\varepsilon_{\rm crit}$ and 
$\Delta \varepsilon$.
Therefore, they are represented by one and the same point in the phase diagram of Fig.~\ref{fig:AHP-phases}, which also by construction lies on the Seidov border line (\ref{seidov}). 
Changing the stiffness of the high-density phase by increasing the value of $\Gamma_3$ moves the border that divides regions for disconnected (D) and absent (A) stable hybrid star branches so that for set 1 our model
lies in the domain A, while for the stiffer sets 2-4 it is in the region D.
The fact that the point representing our parametrizations lies on the Seidov border is merely of academic interest here. 
It indicates that at the onset of the phase transition, if $\Delta \varepsilon$ were just a little smaller, there would be a small connected branch of hybrid stars with a tiny quark matter core  before the onset of the unstable branch, which makes out the difference that one has now B and  C instead of D and A. If one would disregard this tiny "academic" detail, then the regions just below the Seidov line should also be called D and A, respectively.       

\begin{figure}[!htpb]
\includegraphics[width=0.7\textwidth, angle=0]{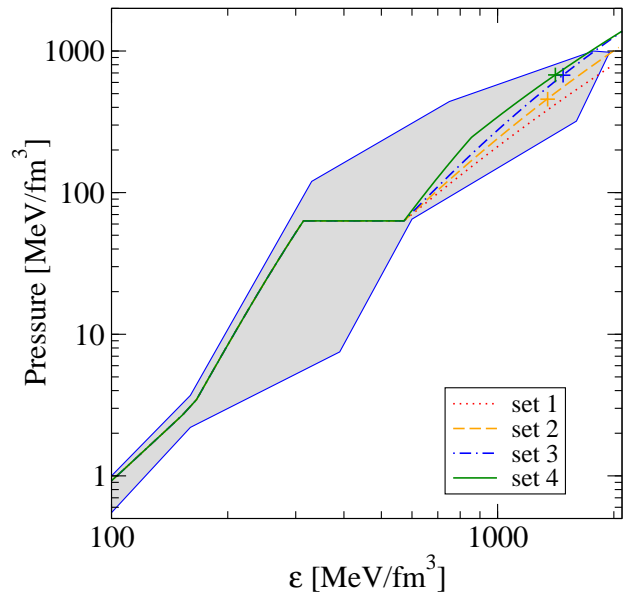}\\
\caption{(Color online) All multi-polytrope EoS for sets 1-4 from Tab.~\ref{param-123} fall within the region given in Hebeler et al.~\cite{Hebeler:2013nza} for the case supporting a $1.97~M_{\odot}$ NS
(grey shaded region).
These EoS share the hadronic branch EoS, the onset density and jump in energy density at the transition 
but vary in the stiffness of the high-density phase. The plus symbols denote the pressure and energy density values at the center of the compact star configuration with the maximum mass.
\label{EoS-MP-Hebeler}}
\end{figure}
All the EoS models in this work are causal and fall inside the EoS region that was given in 
Hebeler et al.~\cite{Hebeler:2013nza} for the case supporting a $1.97~M_{\odot}$ NS,
see Fig.~\ref{EoS-MP-Hebeler}.
\section{Conclusions}
In this work we have shown that within the multi-polytrope approach by Hebeler et 
al.~\cite{Hebeler:2013nza} we can obtain high mass twins in the M-R diagram for compact stars. 
We have also shown that going beyond the three-polytrope scheme one can achieve an increase in the
maximum mass on the third family branch of compact hybrid stars.
This feature is of particular interest for scenarios of NS-NS merger events where in the spin-down evolution of the supermassive compact star a phase transition can occur.

The EoS parametrizations presented here obey causality and fall in the constraint region derived in \cite{Hebeler:2013nza} for the condition that a mass of $1.97~M_\odot$ has to be reached. 

In concluding, we mention that the multi-polytrope approach can of course not replace a realistic EoS. 
While we have shown that extending the three-polytrope scheme to four polytropes gives already a significant improvement of the maximum mass for the hybrid star configurations on the third family branch solutions, the most promising strategy shall be to employ 
microscopic approaches to high-density quark matter like, e.g., the NJL model with higher order repulsive interactions~\cite{Benic:2014jia} or the relativistic string-flip model~\cite{Kaltenborn:2017hus} do not have this problem of multi-polytropes. 
The multi-polytrope scheme can give, however, interesting heuristic guidance along this path.

Finally, we would like to emphasize the importance of the HMT phenomenon detection and it's relation to the QCD critical point~\cite{Blaschke:2013ana,Alvarez-Castillo:2015xfa}. 
If a first order phase transition exists in the QCD diagram it should feature a critical
end up that borders the crossover region. 
Thus, this transition shall extend into isospin asymmetric matter covering the low temperature region where compact star matter is located. 
As we have presented in this work, if the strength of this transition
satisfies the Seidov conditions and quark matter is sufficiently stiff, the high mass twins phenomenon shall occur. Potential detection relies on accurate radius measurements which, for the models presented here, should be capable of resolving a $2-3 $ km difference with confidence.
Bayesian techniques that utilize astronomical measurements can provide model parameter estimation useful to probe the compact star equation of state~\cite{Alvarez-Castillo:2016oln}.

\subsection*{Acknowledgements}
We acknowledge discussions with Jim Lattimer and Jochen Wambach. 
D.B. is grateful for support of his participation at the CERN Theory Institute 
"From quarks to gravitational waves", where these results have first been presented. 
D.B. was supported in part by the MEPhI Academic Excellence Project under 
contract No. 02.a03.21.0005.
This work was supported by the Polish NCN under grant No.
UMO-2014/13/B/ST9/02621.

\end{document}